\documentclass[aps,prb,twocolumn,citeautoscript,superscriptaddress]{revtex4}
\usepackage{amsmath}
\usepackage{graphicx}
\usepackage{dcolumn}
\usepackage{float}
\usepackage{color}
\usepackage{multirow}
\usepackage{amssymb}
\usepackage{microtype}
\usepackage{booktabs}

\begin{document}

\title{ Microscopic evidence for anisotropic multigap superconductivity \\ in the CsV$_3$Sb$_5$ kagome superconductor}

\author{Ritu Gupta\footnotemark[2]}
 \email{ritu.gupta@psi.ch}
 
  \affiliation{Laboratory for Muon Spin Spectroscopy, Paul Scherrer Institute, CH-5232 Villigen PSI, Switzerland}

 \author{Debarchan Das}
\thanks{These authors contributed equally}
 \affiliation{Laboratory for Muon Spin Spectroscopy, Paul Scherrer Institute, CH-5232 Villigen PSI, Switzerland}
 
  \author{Charles Hillis Mielke III}
\thanks{These authors contributed equally}
 \affiliation{Laboratory for Muon Spin Spectroscopy, Paul Scherrer Institute, CH-5232 Villigen PSI, Switzerland}

   \author{Zurab Guguchia}
 \affiliation{Laboratory for Muon Spin Spectroscopy, Paul Scherrer Institute, CH-5232 Villigen PSI, Switzerland}
 
   \author{Toni~Shiroka}
 \affiliation{Laboratory for Muon Spin Spectroscopy, Paul Scherrer Institute, CH-5232 Villigen PSI, Switzerland}
  \affiliation{Laboratorium für Festk{\"o}rperphysik, ETH Z{\"u}rich, Z{\"u}rich, CH-8093, Switzerland}
  
    \author{Christopher Baines}
 \affiliation{Laboratory for Muon Spin Spectroscopy, Paul Scherrer Institute, CH-5232 Villigen PSI, Switzerland}
 
 \author{Marek Bartkowiak}
 \affiliation{Laboratory for Neutron and Muon Instrumentation, Paul Scherrer Institute, CH-5232 Villigen PSI, Switzerland}
 
    \author{Hubertus Luetkens}
 \affiliation{Laboratory for Muon Spin Spectroscopy, Paul Scherrer Institute, CH-5232 Villigen PSI, Switzerland}

\author{Rustem Khasanov}
 \email{rustem.khasanov@psi.ch}
 \affiliation{Laboratory for Muon Spin Spectroscopy, Paul Scherrer Institute, CH-5232 Villigen PSI, Switzerland}
\author{Qiangwei Yin}
\affiliation{Department of Physics and Beijing Key Laboratory of Opto-electronic Functional Materials \& Micro-nano Devices, Renmin University of China, Beijing 100872, China}
\author{Zhijun Tu}
\affiliation{Department of Physics and Beijing Key Laboratory of Opto-electronic Functional Materials \& Micro-nano Devices, Renmin University of China, Beijing 100872, China}
\author{Chunsheng Gong}
\affiliation{Department of Physics and Beijing Key Laboratory of Opto-electronic Functional Materials \& Micro-nano Devices, Renmin University of China, Beijing 100872, China}
\author{Hechang Lei}
\email{hlei@ruc.edu.cn}
\affiliation{Department of Physics and Beijing Key Laboratory of Opto-electronic Functional Materials \& Micro-nano Devices, Renmin University of China, Beijing 100872, China}

\date{\today}

\begin{abstract}
The recently discovered kagome superconductor CsV$_3$Sb$_5$ ($T_c \simeq 2.5$\,K) has been found to host charge order as well as a non-trivial band topology, encompassing multiple Dirac points and probable surface states. Such a complex and phenomenologically rich system 
is, therefore, an ideal playground for observing unusual electronic phases.
Here, we report on microscopic studies of its anisotropic superconducting properties 
by means of transverse-field muon spin rotation ($\mu$SR) experiments. 
The temperature dependences of the in-plane and out-of-plane components of the magnetic penetration depth $\lambda_{ab}^{-2}(T)$ and $\lambda_{c}^{-2}(T)$ indicate that the superconducting order parameter exhibits 
a two-gap ($s+s$)-wave symmetry, reflecting the multiple Fermi 
surfaces of CsV$_3$Sb$_5$.
The multiband nature of its superconductivity is further validated by the different temperature dependences of the anisotropic magnetic penetration depth $\gamma_\lambda(T)$ and upper critical field $\gamma_{\rm B_{c2}}(T)$, both in close analogy with the well known two-gap superconductor MgB$_2$. Remarkably, the high value of the $T_c/\lambda^{-2}(0)$ ratio in both field orientations strongly suggests the unconventional nature of superconductivity. The relaxation rates obtained from zero field $\mu$SR experiments do not show noticeable change across the superconducting transition, indicating that superconductivity does not break time reversal symmetry.   
\end{abstract}

\maketitle

\section{INTRODUCTION}
The kagome lattice materials, consisting of a two-dimensional lattice of corner-sharing triangles, have drawn considerable attention in recent years \cite{JXYin1,GuguchiaCSS,Mazin,LYe}. Their electronic structure is characterized by a dispersionless flat band, whose origin lies in the innate kinetic frustration of the kagome geometry, and a pair of Dirac points \cite{CaoPablo}. Such flat bands, with a high density of electronic states, are generally perceived to quench the kinetic energy and to induce 
correlated electronic phases when found close to the Fermi level \cite{Hofmann2020,Nunes}, as illustrated 
by the recently discovered superconducting twisted bilayer graphene   \cite{CaoPablo}. The inherent geometrical frustration of kagome systems can be employed 
to carefully tune their properties, thus aiding in the search of 
superconductors (SC) with non-phonon mediated pairing mechanisms \cite{CMielke}. A recent example of a kagome superconductor with unconventional 
coupling is 
LaRu$_{3}$Si$_{2}$  \cite{CMielke}. Here, the correlation effects from the kagome flat band, the van Hove points on the kagome lattice, and the high density of states from the narrow electronic bands were proposed as key factors for achieving a  relatively high transition temperature $T_c~\simeq~7$\,K.

Following the recent discovery of the $A$V$_{3}$Sb$_{5}$ ($A$~=~K,~Rb,~Cs) family of kagome materials \cite{BOrtiz1}, a slew of interesting and exotic effects have been observed: giant anomalous Hall conductivity \cite{SYang,FYu,HZhao}, magneto-quantum oscillations \cite{SYang,YFu}, topological charge order \cite{YJiang,MDenner,NShumiya,Wang2021,MKang,HLi}, orbital order \cite{DSong}, and superconductivity \cite{BOrtiz2,BOrtiz3,QYin,ZLiu}. Featuring a kagome network of vanadium atoms interwoven with a simple hexagonal antimony net, the normal state of CsV$_{3}$Sb$_{5}$ was described as a nonmagnetic $Z_2$ topological metal \cite{BOrtiz2, BOrtiz3}. Furthermore, the observation of CDW order in the normal state of all members of $A$V$_{3}$Sb$_{5}$ kagome family has generated 
significant theoretical and experimental interest. Namely, topological chiral charge order has been reported in {\it A}V$_3$Sb$_5$ ($A$ = K, Rb, or Cs) \cite{YJiang,NShumiya}. In KV$_{3}$Sb$_{5}$, direct evidence for time-reversal symmetry breaking by the charge order was demonstrated using 
muon spin rotation \cite{Mielke2021}. 

Regarding superconductivity, a strong diversity in the SC gap symmetry is reported.  Proximity-induced spin-triplet pairing was suggested for Nb-K$_{1-x}$V$_{3}$Sb$_{5}$ \cite{YWang}. In CsV$_{3}$Sb$_{5}$, multiband superconductivity with sign-preserving order parameter was reported in Refs.~\cite{HXu,KNakayama,Duan}, while a roton-pair density wave, as well as anisotropic- and nodal superconducting pairing were suggested in Refs.~\cite{HChen,SNi,CZhao} Finally, reentrant superconductivity and double SC domes were found under pressure \cite{ZZhang,XChen,KChen}. From a  theoretical perspective, several scenarios for electronically mediated, unconventional superconductivity have been discussed \cite{XWu}. The $A$V$_3$Sb$_5$ electronic bands exhibit van Hove singularities close to the Fermi energy --- an electronic structural motif shared with other systems, such as the cuprate superconductors or Sr$_2$RuO$_4$. A particular feature of the kagome lattice, however, is a sublattice interference mechanism \cite{Kiesel}, by which the Bloch states near each van Hove point are supported on a distinct sublattice. This promotes the relevance of long-range interactions and unconventional pairing states. 

\begin{figure*}[htb!]
\includegraphics[width=1.0\linewidth]{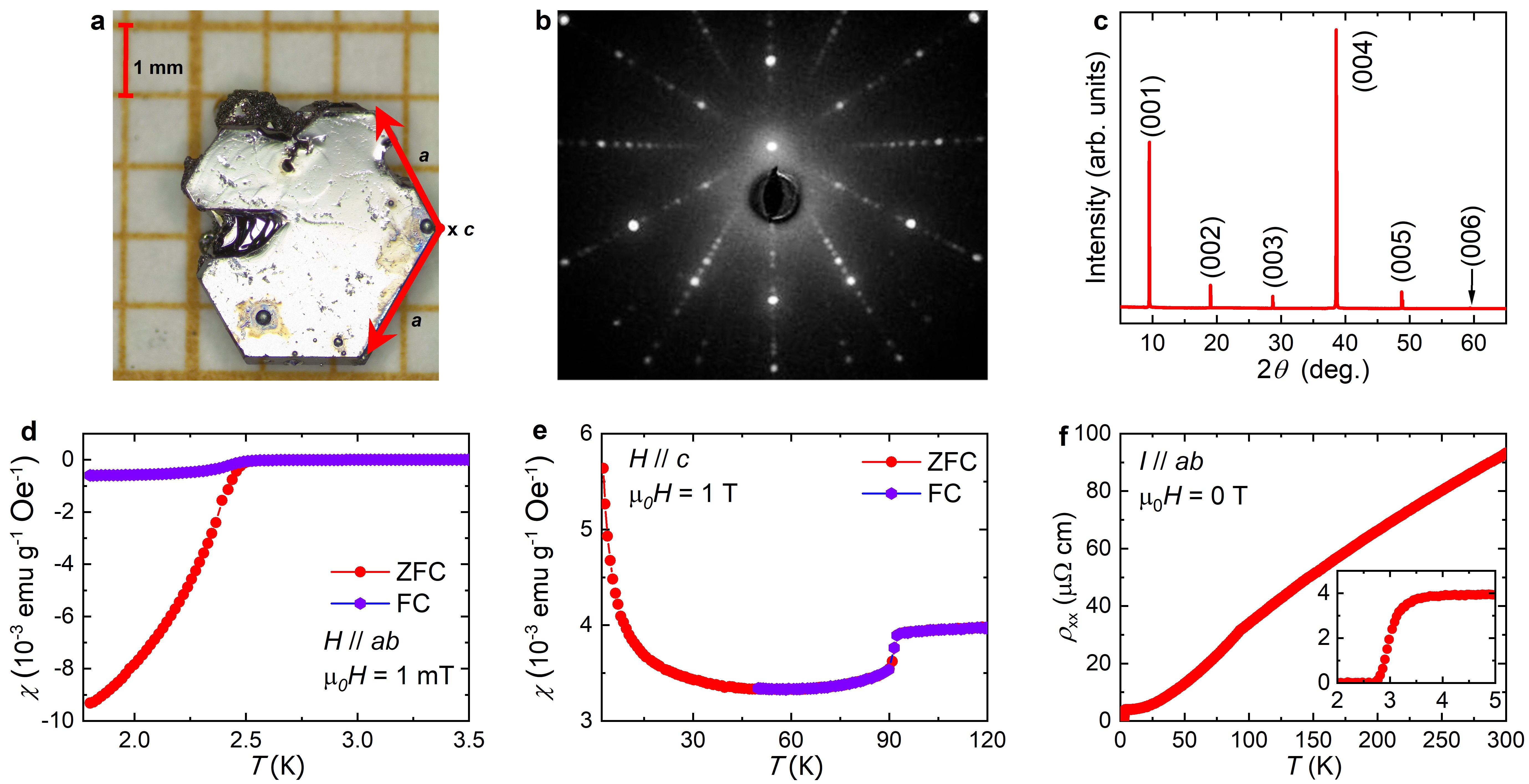}
\caption{(Color online) Characterization of CsV$_3$Sb$_5$ single crystals. (a) Optical microscope photograph of a single crystal with the crystallographic axes highlighted by arrows. Notice the obvious hexagonal symmetry. (b) Laue x-ray diffraction image of the single crystal along the 001 direction. (c) X-ray diffraction pattern obtained from a single crystal sample. (d) Temperature dependence of magnetic susceptibility in an applied field of 1~mT along the $ab$ plane highlighting the superconducting transition. (e) Temperature dependence of magnetic susceptibility in an applied field of 1~T along the $c$ axis emphasizing the CDW transition at 95\,K. (f) Electrical resistivity as a function of temperature under zero applied field.}
\label{fig:fig1}
\end{figure*}
To explore unconventional aspects of superconductivity in CsV$_{3}$Sb$_{5}$, it is critical to measure the superconducting order parameter on the microscopic level through measurements of the bulk properties. Thus, we focus on muon spin rotation/relaxation (${\mu}$SR) measurements of the magnetic penetration depth \cite{Sonier} $\lambda$ in CsV$_{3}$Sb$_{5}$. $\lambda$ is one of the fundamental parameters of a superconductor, since it is related to the superfluid density $n_{s}$ via:
\begin{equation}
 1/{\lambda}^{2} = \mu_{0}e^{2}n_{s}/m^{*},  
\end{equation}
where $m^{*}$ is the effective mass. Most importantly, the temperature dependence of ${\lambda}$ is particularly sensitive to the topology of the SC gap: while in a fully gapped superconductor, $\Delta\lambda^{-2}\left(T\right)\equiv\lambda^{-2}\left(0\right)-\lambda^{-2}\left(T\right)$ vanishes exponentially at low $T$, in a nodal SC it shows a linear $T$-dependence.

\section{Results}
To determine sample purity, X-ray diffraction experiments were performed on flux-grown crystals. The powder diffraction pattern of ground crystals can be well fitted using the structure of CsV$_3$Sb$_5$ 
and the fitted lattice parameters are $a$ = 5.50552(2) \AA~and $c$ = 9.32865(3) \AA , close to the previous results \cite{BOrtiz3}. Additionally, to test the single crystallinity of the samples and determine the orientation for the $\mu$SR experiments, X-ray Laue diffraction was performed on the single crystal shown in Fig.~\ref{fig:fig1}~(a), whose diffraction pattern is displayed in Fig.~\ref{fig:fig1}~(c). The crystal was easily aligned; the hexagonal symmetry of the $ab$-plane is clearly visible from the single crystal, and the crystals grow with the $c$-axis aligned along the thin direction of the crystal. The diffraction pattern collected in Fig.~\ref{fig:fig1}~(c) was analyzed with the OrientExpress program \cite{OrientExpress} and the orientation was confirmed to be  along the crystallographic $c$-axis. 
  
The superconductivity of the samples was confirmed by magnetization [Fig.~\ref{fig:fig1}~(d)] and resistivity [Fig.~\ref{fig:fig1}~(f)] experiments, which show a diamagnetic shift in the sample concurrent with the onset of zero-resistivity at $T_{\rm c} \simeq 2.7 $~K which is slightly higher than the compared to the  $T_{\rm c}$ (= 2.5~K) value obtained from magnetization measurements. This is most likely due to very tiny filamentary superconducting channels. The onset of charge order is visible in the magnetization measurements in Fig.~\ref{fig:fig1}~(e), corresponding to the anomaly at $T_{\rm co}~\approx$~95~K. There is also a slight change in slope visible in the resistivity data presented in Fig.~\ref{fig:fig1}~(f), which occurs at the onset of the charge-ordered state.


\begin{figure*}[htb!]
\includegraphics[width=1.0\linewidth]{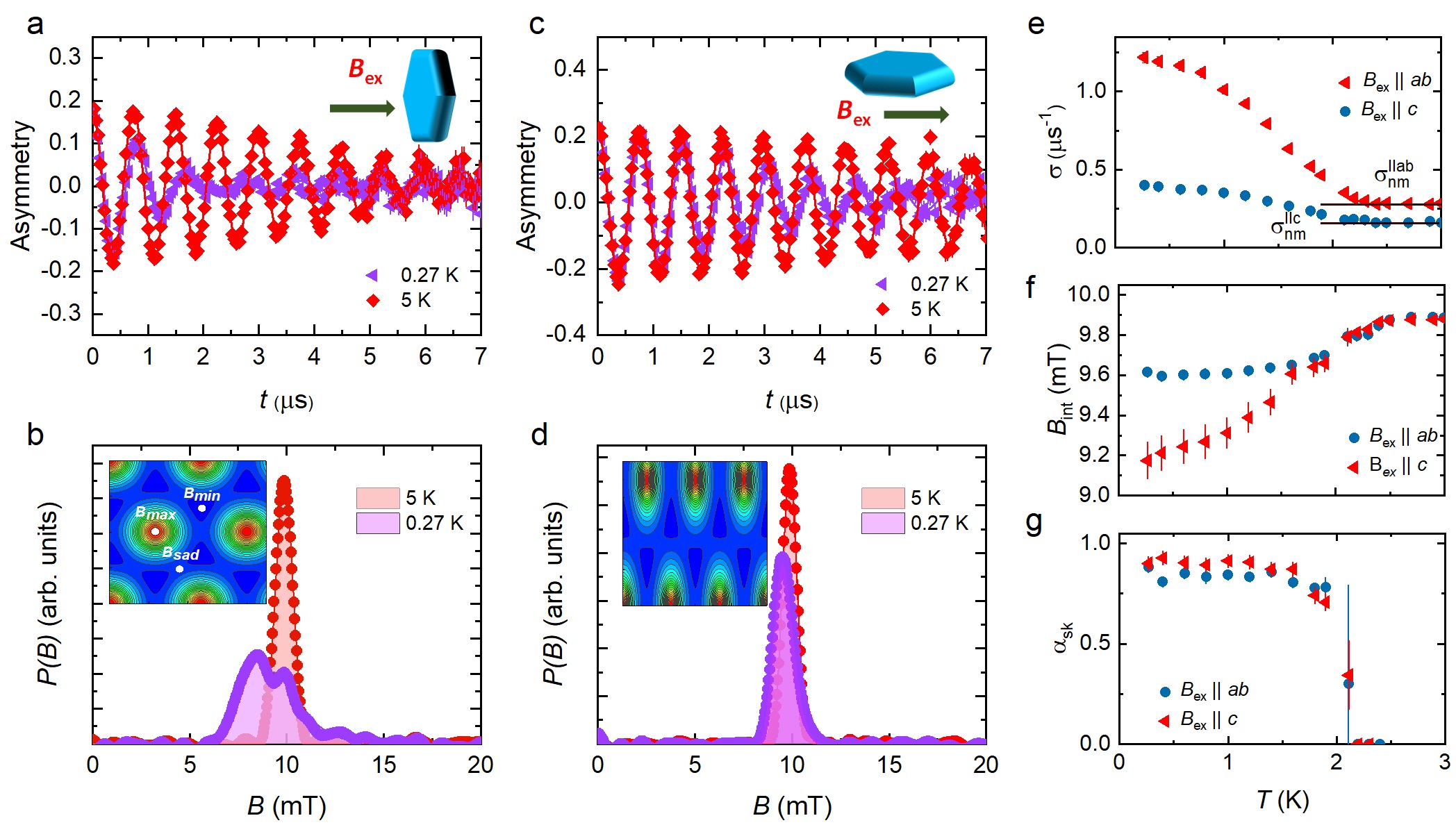}
\caption{(Color online) (a) and (c) TF-${\mu}$SR spectra collected above and below $T_{c}$ for CsV$_3$Sb$_5$ with $B_{\rm ext}||c$ and $B_{\rm ext}||ab$, respectively. The fast damping of the signal in the superconducting state reflects the inhomogeneous field distribution sensed by the muon ensemble as a result of vortex formation. Solid lines through the data points are fits 
using Eq.~(5) in the Suppl. Inset: Scheme of the setup for $\mu$SR experiments on a  single crystalline CsV$_3$Sb$_5$ sample. (b) and (d) show the Fourier spectra obtained by fast Fourier transformation of the spectra in (a) and (c), respectively. Inset: Contour plot of the field variation in the triangular lattice. $B_{\rm min}$, $B_{\rm max}$, and $B_{\rm sad}$ are respectively the minimum, maximum, and saddle field points. (e) Temperature variation of the square-root of the second moment of field distribution $\sigma(T)$, as measured for the two applied field directions. The horizontal lines 
denote the relaxation due to nuclear magnetic moments along the two directions. (f) Internal magnetic field as a function of temperature in the superconducting state. A clear diamagnetic shift, an immanent feature of type-II superconductors, can be seen. (g) Temperature variation of the skewness parameter, $\alpha_{\rm sk} = M_3^{1/3}/M_2^{1/2}$, for the $B_{\rm ext}||c$ and $B_{\rm ext}||ab$ set of experiments.}
 \label{fig:figure2}
\end{figure*}


  Two sets of TF-$\mu$SR experiments were carried out in the field-cooled state, with the external magnetic field applied parallel to the $c-$axis, and parallel to the $ab$ (kagome) plane. 
  In both cases the  muon spin was perpendicular to the applied field. Note that, for an applied field parallel to the $c$-axis, the screening currents around the flux-line cores flow in the $ab$-plane. This allows us to determine the so-called in-plane component of the magnetic penetration depth $\lambda_{ab}$. The TF-$\mu$SR time-spectra collected with an external field $B_{\rm ext}~=~10$~mT applied parallel to the $c$-axis above ($T = 5$~K) and well below ($T\simeq0.27$~K) the superconducting transition temperature $T_{\rm c}\simeq~2.5$~K are shown in Fig.~\ref{fig:figure2}~(a). The corresponding Fourier transforms of the TF-$\mu$SR data, representing the magnetic field distribution $P(B)$, are shown in Fig.~\ref{fig:figure2}~(b). The insets in Fig.~\ref{fig:figure2}~(a) and Fig.~\ref{fig:figure2}~(b), respectively, represent the geometry of the experiment and the schematic distribution of the magnetic fields within the isotropic flux-line lattice (FLL) with the two components of the magnetic penetration depth, namely $\lambda_a$ and $\lambda_b$ being equal: $\lambda_a=\lambda_b=\lambda_{ab}$. The TF-$\mu$SR time-spectra and the corresponding Fourier transforms collected with $B_{\rm ext} = 10$\,mT applied along the kagome plane are presented in Figs.~2(c) and (d), respectively. With the field applied along the $ab$-plane, the screening currents around the vortex cores flow along the $ab$-plane and $c$-axis, thus implying that in a set of experiments with $B \| ab$, $\lambda_{ab,c}$ can be determined. Note that, due to the anisotropy, $\lambda_{c}$ is longer than $\lambda_{ab}$, which leads to an elongation of the vortex lattice along the $c$ direction [see inset in Fig.~\ref{fig:figure2}~(d)].

The formation of the flux-line lattice (FLL) in the superconductor leads to a nonuniform magnetic field distribution between the vortices [see insets in Figs.~\ref{fig:figure2}~(b) and (d)]. The strong damping of the TF-$\mu$SR time-spectra [Figs.~\ref{fig:figure2}~(a) and (c)] and the corresponding broadening of the Fourier transform [Figs.~\ref{fig:figure2}~(b) and (d)] represent exactly this effect. Note that the measured distribution of the magnetic fields in the superconducting state becomes asymmetric, as expected for a well-arranged FLL. All the characteristic features, as e.g., the cutoff at low fields ($B_{\rm min}$), the peak due the saddle point between two adjacent vortices ($B_{\rm sad}$), and the long tail towards high fields, related to the regions around the vortex core ($B_{\rm max}$), are clearly visible in both field orientations. The locations of $B_{\rm min}$, $B_{\rm sad}$, and $B_{\rm max}$ are shown in the contour plot in the inset of Fig.~\ref{fig:figure2}~(b). To account for the field distribution $P(B)$, the time-domain spectra were analyzed using a skewed Gaussian (SKG) function, which represents the simplest distribution accounting for the asymmetric lineshape (see Suppl.\ Mater.\ for a detailed description of the function).

 The parameters obtained from the fits are presented in Fig.~\ref{fig:figure2}~(e-g). Fig.~\ref{fig:figure2}~(e) shows the temperature dependence of the square root of the second moment $M_2$ (see Suppl.\ Mater.\ for detailed calculations) which corresponds to the total depolarization rate $\sigma$ for two field orientations. Below $T_{c}$, the relaxation rate $\sigma$ starts to increase from its normal-state value due to the formation of the FLL and saturates at low temperatures. The normal-state muon depolarization rate is mostly due to the nuclear magnetic moments and, for CsV$_3$Sb$_5$, it has different values for the two field orientations, $\sigma_{\rm nm}^{||c}$~=~0.165(3)~$\mu$s$^{-1}$ and $\sigma_{\rm nm}^{||ab}$~=~0.287(3)~$\mu$s$^{-1}$ [shown by horizontal lines in Fig.~\ref{fig:figure2}~(e)]. As shown in Fig.~\ref{fig:figure2}~(f), for both field orientations the first moment --- which represents the internal field ($B_{\rm int}$) --- shows a clear  diamagnetic shift  below $T_{\rm c}$, as expected for a type-II superconductor. Note that the $T_{\rm c}$ value estimated from $\mu$SR experiments agrees with that determined from magnetization measurement, so that $T_{\rm c}^{{\mu}{\rm SR}}=T_{\rm c}^{{\rm \chi}}\simeq 2.5$~K.

The asymmetric line shape of the field distribution $P(B)$ is characterized by the third moment ($M_3$) of the field distribution (see Suppl.\ Mater.\ for the calculations) and has three characteristic fields: $B_{\rm min}$, $B_{\rm max}$, and $B_{\rm sad}$.
More accurately, the asymmetry of the line shape is described by its skewness parameter $\alpha_{\rm sk} = (M_3^{1/3}/M_2^{1/2})$,
%
which assumes a value of 1.2 for a perfectly arranged triangular vortex lattice \cite{Aegerter}. Distortions or even melting of the vortex lattice structure, which may be caused by variations of temperature or magnetic field, are strongly reflected in $\alpha_{\rm sk}$\cite{Lee, Blasius, Heron}.  

Figure~2(g) shows the temperature evolution of $\alpha_{\rm sk}^{||c}$ (for $B_{\rm ext}||c$) and $\alpha_{\rm sk}^{||ab}$ (for $B_{\rm ext}||ab$) for the kagome superconductor CsV$_3$Sb$_5$. Notably, in both directions, $\alpha_{\rm sk} (T)$ remains independent of temperature for $T \lesssim T_c$, with a constant value of $\simeq0.8$ for $\alpha_{\rm sk}^{||ab}$ and 0.9 for $\alpha_{\rm sk}^{||c}$, respectively. We note that near the superconducting transition temperature the $\mu$SR response could be well fitted by the single Gaussian line (i.e. the reduced $\chi^2$ of SKG and single Gaussian fits become almost equal). Since for the symmetric $P(B)$ distribution $\alpha_{\rm sk}$ stays exactly at zero, this leads to the sudden change of $\alpha_{\rm sk}$ at $T\sim 2.2~K$, i.e. $\simeq 0.3~$K below $T_{\rm c}$.  This observation suggests that very close to $T_{\rm c}$, the FLL is slightly distorted but does not disturb the determination of the temperature evolution of the superfluid density along different crystallographic directions (as we show later), which is the main goal of the present study. 




We estimate the superconducting contribution to the depolarization rate, $\sigma_{\rm sc}$, by quadratically subtracting the temperature-independent nuclear magnetic moment contribution $\sigma_{\rm nm}$ (obtained above $T_c$) from the the total depolarization rate $\sigma$ (i.e., $\sigma_{\rm sc}^2 = \sigma^2-\sigma_{\rm nm}^2$).
$\sigma\textsubscript{sc}$ can be expressed as a function of the reduced field $b$ = $\frac{B}{B_{c2}}$ ($B_{c2}$ being the upper critical field) and the Ginzburg-Landau coefficient $\kappa$ by the relation developed by Brandt \cite{Brandt_PRB_1988,Brandt_PRB_2003}
\begin{equation}
\sigma_{sc}[\mu \mathrm{s}^{-1}]\approx 4.83\times 10^{4}(1-b)[1+1.21(1-\sqrt{b})^3]\lambda^{-2}[\mathrm{nm}^{-2}],
\end{equation}
where $\lambda$ is the magnetic penetration depth.
Thus, we obtain the temperature dependence of $\lambda_{ab}^{-2}$ (for $B_{\rm ext}\parallel c$) and $\lambda_{ab,c}^{-2}$ (for $B_{\rm ext}\parallel ab$).
In the case of an anisotropic superconductor, the magnetic penetration depth is also anisotropic. In the present case, by considering an anisotropic effective mass tensor, $\lambda_{ab}^{-2}$ and 
$\lambda_{ab,c}^{-2}$, $\lambda_{c}^{-2}$ can be estimated (see Suppl.\ Mater.\ for the detailed analysis). In this way, we can directly compare the anisotropy of the magnetic penetration depth.  Figures~3~(a) and (b) show the temperature evolution of $\lambda_{ab}^{-2}$ and $\lambda_{c}^{-2}$, respectively.

\begin{figure*}[htb!]
\includegraphics[width=1.0\linewidth]{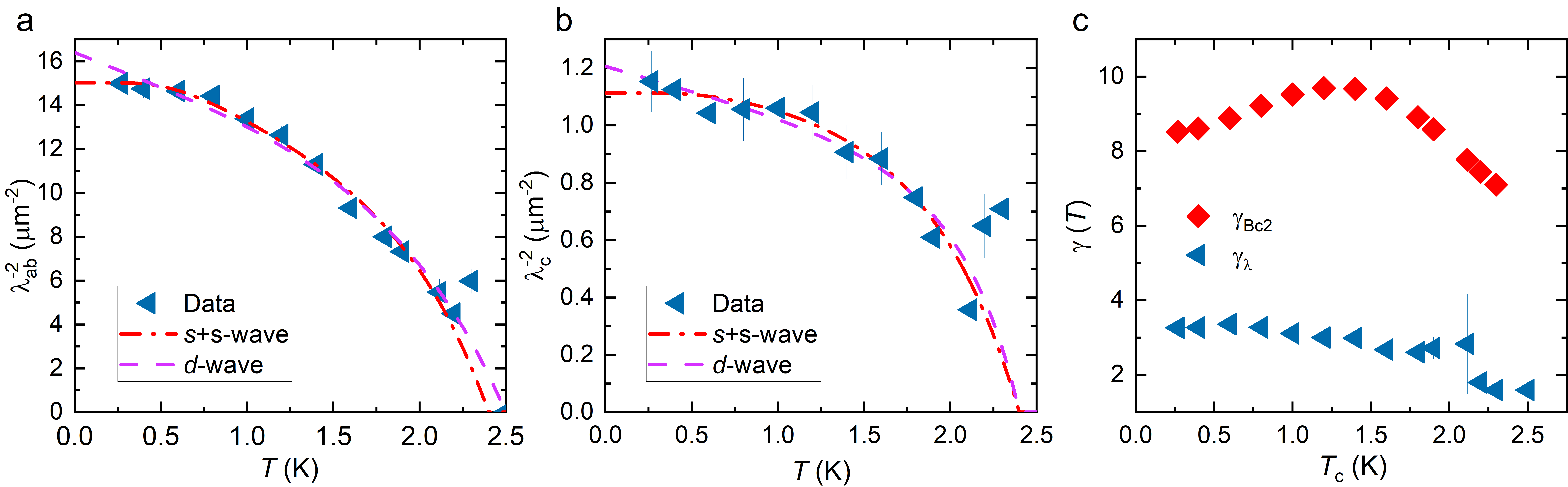}
\caption{(Color online) (a) and (b) Temperature variations of the in-plane and out-of-plane component of the inverse squared magnetic penetration depths, $\lambda_{ab}^{-2}(T)$ and $\lambda_{c}^{-2}(T)$. Data are fitted with a two gap $(s+s)$-wave, and a nodal $d$-wave model, as described by the phenomenological $\alpha$-model. (c) Magnetic penetration depth anisotropy ($\gamma_\lambda$) and upper critical field anisotropy ($\gamma_{B_{c2}}$) as a function of temperature. $\gamma_{B_{c2}}(T)$ is obtained by considering the SC transition temperature $T_c$ as the temperature where resistivity approaches zero.}
\label{fig:figure3}
\end{figure*}

As mentioned in the introduction, there are only a few studies involving different experimental techniques to address the superconducting gap structure of the CsV$_3$Sb$_5$ kagome superconductor. However, there is no consensus among them. To determine whether the superconducting gap structure of this compound is of single-gap, multigap, or even of nodal nature, we analyzed the temperature dependence of magnetic penetration depth. We analyzed $\lambda(T)$ data using the following expression:
\begin{equation}
\frac{\lambda^{-2}(T,\Delta_0)}{\lambda^{-2}(0,\Delta_0)}=1+\frac{1}{\pi}\int_{0}^{2\pi}\!\!\int_{\Delta (T, \phi)}^{\infty}\frac{\partial f}{\partial E } \frac{E\,\mathrm{d}E\,\mathrm{d}\phi}{\sqrt{E^2-\Delta (T, \phi)^2}},
\end{equation}
where $f=(1+E/k_\mathrm{B}T)^{-1}$ represents the Fermi-distribution function. The temperature- and angle-dependent gap function is described by $\Delta (T, \phi)=\Delta_0\delta(T/T_c)g(\phi)$, where ${\Delta}_{0}$ is the maximum gap value at $T=0$, while the temperature dependence of the gap function is\cite{Rustem_AuBe, Carrington} \mbox{$\delta(T/T_c)=\tanh\{1.821[1.018(T_c/T-1)^{0.51}]\}$}. Here, $g({\varphi}$) corresponds to the angular dependence of the gap and  takes a value of 1 for $s$-wave and ${\mid}\cos(2{\varphi}){\mid}$ for $d$-wave gap symmetry. Motivated by recent studies reporting a nodal-gap structure in CsV$_3$Sb$_5$, evidenced by a non-zero value of the residual linear term of its thermal conductivity at zero field together with its rapid increase with field, we tried to fit the data using a $d$-wave model\cite{CZhao}. However, as shown in Figs.~\ref{fig:figure3}~(a) and (b), a $d$-wave model cannot describe the data well. On the contrary, recent tunneling experiments\cite{Xu_CVS}, tunnel-diode oscillator (TDO) based results, along with specific heat measurements conjointly indicate a multiband nature of superconductivity (SC) in CsV$_3$Sb$_5$ \cite{Duan}. Thus, we proceeded
to fit the $\lambda_c^{-2} (T)$ and $\lambda_{ab}^{-2} (T)$ data simultaneously with a two-gap scenario using a weighted sum:
\begin{equation}
\frac{\lambda\textsuperscript{-2}(T)}{\lambda\textsuperscript{-2}(0)}= x\frac{\lambda\textsuperscript{-2}(T,\Delta_{0,1})}{\lambda\textsuperscript{-2}(0,\Delta_{0,1})}+(1-x)\frac{\lambda\textsuperscript{-2}(T,\Delta_{0,2})}{\lambda\textsuperscript{-2}(0,\Delta_{0,2})}.
\end{equation}
%
\noindent Here $x$ is the weight associated with the larger gap and $\Delta_{0,i}$ ($i$ = 1, 2 are the band indices) are the gaps related to the first and second band.

The fit of $\lambda_{ab}(T)$ and $\lambda_{c}(T)$ with a two-gap $(s+s)$-wave model was performed by assuming similar gap values ($\Delta_{1(2),ab} = \Delta_{1(2),c}$), but different weighting factors for the two directions ($x_{ab} \neq x_c$). It is evident that the two-gap model follows the experimental data very well [Figs.~\ref{fig:figure3}~(a) and (b)]. 
Table~I summarizes the different superconducting parameters as obtained from the fits. Note that the TF-$\mu$SR measurements are well described by gap values which are very close to those obtained from tunneling measurements ($\Delta_1$ = 0.57 meV, $\Delta_2$ = 0.3 meV, $\Delta_3$ = 0.45 meV) \cite{Xu_CVS}. On the other hand, as shown in Figs.~\ref{fig:figure3}~(a) and (b), a $d$-wave model does not describe the data well, thus ruling out the possibility of nodes in the SC gap structure.


Furthermore, we determined the anisotropy of the magnetic penetration depth, $\gamma_{\lambda}$, defined as:
\begin{equation}
\gamma_{\lambda} = \frac{\lambda_c}{\lambda_{ab}} = \frac{\lambda_{ab}^{-2}}{\lambda_{ab,c}^{-2}}.
\label{eq:gamma_lambda}
\end{equation}
By  using the $\lambda_{ab}^{-2}(T)$ and $\lambda_{c}^{-2}(T)$ values, we obtain the temperature dependence of the magnetic penetration depth anisotropy as presented in Fig.~\ref{fig:figure3}~(c). According to the phenomenological Ginzburg-Landau theory for uniaxial anisotropic superconductors, the various anisotropies, such as the magnetic penetration depth anisotropy $\gamma_\lambda$ and the upper critical field anisotropy $\gamma_{B_{c2}}$, can be accounted for by a common parameter:\cite{Kogan_PRB_1981, Tinkham_book_1975}
\begin{equation}
\gamma_{\lambda}=\frac{\lambda_c}{\lambda_{ab}}=\sqrt{\frac{m_c^*}{m_{ab}^*}}=\gamma_{B_{\rm c2}}=\frac{B_{\rm c2}^{\parallel ab}}{B_{\rm c2}^{\parallel c}}=\frac{\xi_{ab}}{\xi_c}.
\end{equation}
In the above equation, $\xi$ represents the coherence length. In order to compare $\gamma_{\lambda}$ and $\gamma_{B_{\rm c2}}$, we analyzed the electrical transport data in the presence of various applied fields and estimated the temperature dependence of the upper critical field anisotropy $\gamma_{B_{c2}}(T)$~=~$B_{c2,ab}(T)$/$B_{c2,c}(T)$, where $B_{c2,ab}(T)$ and $B_{c2,c}(T)$ are the upper critical fields corresponding to zero values of resistivity for $B \parallel {ab}$ and $B \parallel c$ (see Fig.~S1 in Suppl.\  Mater.). Figure~\ref{fig:figure3}~(c) depicts also the temperature dependence of $\gamma_{B_{c2}}$. Interestingly, $\gamma_{\lambda}(T)$ changes slightly from $\gamma_{\lambda}\simeq 1.6 $ close to $T_{\rm c}$ to $\gamma_{\lambda}\simeq  3.3$ close to $T=0$ K. Conversely, $\gamma_{B_{c2}}(T)$ varies strongly from $\gamma_{B_{c2}}\simeq 7.1 $ close to $T_{\rm c}$ to $\gamma_{B_{c2}}\simeq  8.5$ when $T_c\simeq$ 0.3 K.   
%


  \begin{figure*}[htb!]
\includegraphics[width=1.0\linewidth]{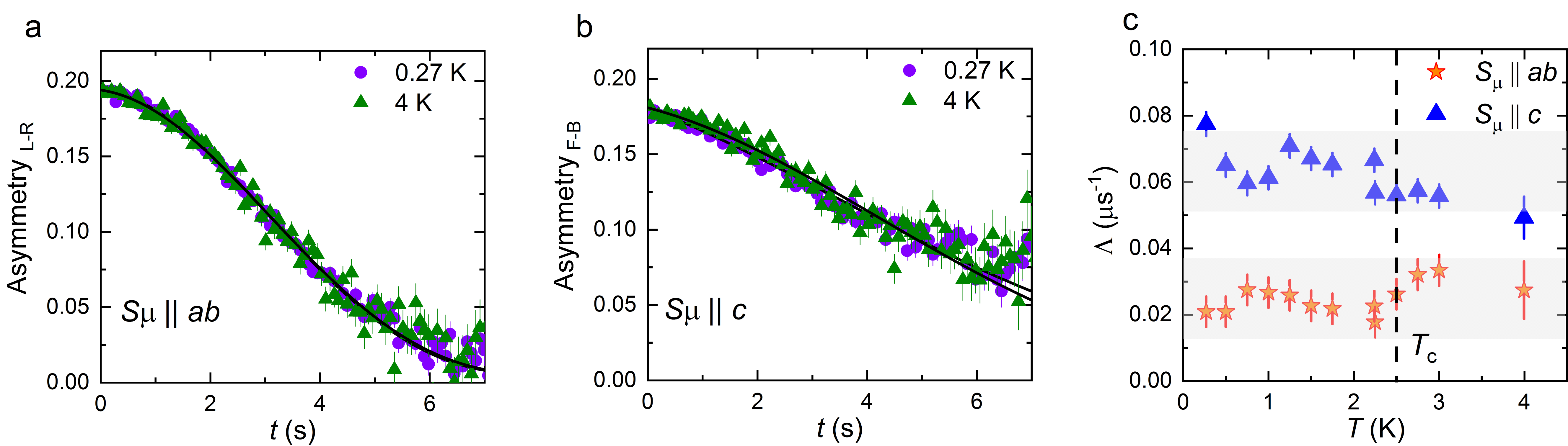}
\caption{(Color online) (a) and (b) ZF-$\mu$SR spectra above and below $T_c$ collected in zero field with the initial muon spin polarization being $S_{\mu}(0)|| ab$ and $S_{\mu}(0)|| c$, respectively. The solid lines through the spectra are fits as described in the text. (c) The resulting ZF exponential relaxation rates $\Lambda (T)$ are almost independent of 
temperature, thus indicating a preserved TRS.}
\label{fig:figure4}
\end{figure*}

\begin{table*}[htb!]
	\centering
	\caption [Supercon para]{Superconducting parameters determined from fits to the temperature dependence of $\lambda(T)$ derived from TF-$\mu$SR experiments, using an ($s+s$)-wave and a $d$-wave  model.}
	\label{table:CVS-superconducting}
	\vskip 3mm
	\addtolength{\tabcolsep}{+5pt}
	\begin{tabular}{c l c c c c c c}
		\toprule
		& Model & $T_{c}$ (K) &  $\Delta_{0, 1}$ (meV)& $\Delta_{0, 2}$ (meV) &$\lambda (0)^{-2}$ ($\mu$m$^{-2}$) & $x$ \\
		\midrule
		$\lambda_{ab}^{-2}$ &($s+s$)-wave & 2.40 & 0.57 & 0.23(3) & 15.0(3) & 0.37(4)\\
        & $d$-wave & 2.50 & 0.62(2)& -- & 16.4(2) & --\\
       	$\lambda_{c}^{-2}$ &($s+s$)-wave & 2.40 &  0.57 & 0.23(3) & 1.11(5) & 0.2(1)\\
       & $d$-wave & 2.40 & 0.82(8)& -- & 1.21(5) & -- \\
		\bottomrule
	\end{tabular}
	\label{SCparameters}
\end{table*}

Finally, to check whether superconductivity breaks time-reversal symmetry in CsV$_3$Sb$_5$, we performed zero-field (ZF) $\mu$SR measurements at different temperatures across $T_{\rm c}$. A time-reversal symmetry-breaking superconducting state is one of the hallmarks of certain unconventional superconductors; in particular, in many Re-based superconductors \cite{Shang2021},  Sr$_2$RuO$_4$ \cite{Luke,Vadim1,Vadim2}, Ba$_{1-x}$K$_x$Fe$_2$As$_2$ \cite{Vadim3}, LaNiC$_2$ \cite{Hillier}, La$_7$Ir$_3$\cite{Barker} etc. Due to the high muon sensitivity to low magnetic fields, $\mu$SR has been the technique of choice for studying it. 
Here, the ZF-$\mu$SR spectra are well described by a damped Gaussian Kubo-Toyabe depolarization function \cite{Hayano1979} (see Supp. Mater. for a detailed description of analysis), which considers the field distribution at the muon site created by both nuclear and electronic moments. The (ZF) $\mu$SR experiments were carried out with the muon spin rotated by $\sim 43^{\circ}$ with respect to the direction of muon momentum. This configuration allows us to use two different detector pairs, namely, Forward(F)- Backward(B) and Left(L)-Right(R), to probe the anisotropic ZF response (see Fig.~S2 in the Suppl.\ Mater.\ for a schematic view of the experimental geometry). Fig.~4~(a) and (b) show the ZF-$\mu$SR spectra collected above and below $T_{c}$. Figure~4~(c) shows the temperature dependence of the electronic relaxation rates for different detectors. The shaded regions represent the statistical scattering of data points. It is evident from Fig.~4~(c) that the relaxation rates do not show any noticeable increase below $T_c$, indicating that superconductivity does not break TRS in CsV$_3$Sb$_5$ within the statistical accuracy, similar to the recent results on the related system KV$_3$Sb$_5$ \cite{Mielke2021}.

\section{Discussion}
The probability field distribution determined experimentally shows a highly asymmetric lineshape, indicative of a well-ordered FLL in the vortex state of the superconductor CsV$_3$Sb$_5$, 
and it remains almost independent of temperature until very close to $T_{\rm c}$. This observation unambiguously suggests that the FLL in CsV$_3$Sb$_5$ is well arranged in the superconducting state and it gets slightly distorted only in the vicinity of $T_{\rm c}$. In general, the change of $\alpha_{\rm sk}$ as a function of magnetic field and temperature is associated with the vortex lattice melting\cite{Lee, Blasius, Rustem_PRL_2008}, and/or a  dimensional crossover from a three-dimensional (3D) to a two-dimensional (2D) type of FLL \cite{Blasius, Aegerter}. Both processes are thermally activated and caused by increased vortex mobility via a loosening of the inter- or intraplanar FLL correlations \cite{Aegerter}. Another possibility involves the rearrangement of the vortex lattice induced by a change of the anisotropy coefficient $\gamma_\lambda =\lambda_c$/$\lambda_{ab}$\cite{Khasanov_PRL_2009}. 
Since CsV$_3$Sb$_5$ has a very small superconducting anisotropy ($\gamma_\lambda \simeq 3$, see below), we can rule out a possible 
vortex-melting scenario. As for the anisotropy-induced FLL rearrangement, the temperature evolution of $\alpha_{\rm sk}$ measured in $B_{\rm ext}\parallel ab$ and $B_{\rm ext}\parallel c$ experiments are expected to be very much different \cite{Khasanov_PRL_2009}, which is also not the case here [see Fig.~3~(c)]. Therefore, we are left with the explanation that close to $T_{\rm c}$, where the broadening of $\mu$SR signal caused by formation of FLL becomes comparable with or even smaller than the relaxation caused by the nuclear magnetic moments [straight lines in Fig.~2~(e)], the shape of $P(B)$ distribution is dominated by the symmetric ‘nuclear’ term, which effectively pushes $\alpha_{\rm sk}$ to zero shortly before the superconducting transition temperature $T_{c}$ is reached. 


Furthermore, a detailed analysis of the $\lambda(T)$ data reveals the presence of two superconducting gaps at the Fermi surface, with gap values of 0.6 and 0.23~meV. This conclusion is in agreement with recent reports involving different experimental techniques\cite{SNi,Duan}. As $\mu$SR is a bulk probe, we conclude that the bulk superconducting gap of this compound consists of two $s$-wave gaps rather than a nodal gap. Another interesting observation is the fact that the $T_{c}/\lambda ^{-2}(0)$ ratio for CsV$_3$Sb$_5$ in both field orientations is comparable to those of high-temperature unconventional superconductors and iron-pnictides \cite{Uemura1, Uemura2}.  Systems with a small $T_{c}/\lambda^{-2}(0)$~$\sim$~0.00025-–0.015 are usually considered to be BCS-like, while large $T_{\rm c}/\lambda^{-2}(0)$ values are expected only in the BEC-like picture and is considered a hallmark feature of unconventional superconductivity. This approach has become a key feature to characterize BCS-like (so-called conventional) and BEC-like superconductors. Remarkably, in CsV$_3$Sb$_5$, $T_{c}/\lambda^{-2}(0)$ is as high as $\sim$~0.2 (for $\lambda_{ab}^{-2}(0)$) - 2.2 ($\lambda_{c}^{-2}(0)$), where the lower limit is comparable to the unconventional transition metal dichalcogenide superconductors \cite{Guguchia_Nat_Com} and the upper limit is close to $T_{c}/\lambda ^{-2}(0)$~$\sim$~4 of hole-doped cuprates \cite{Uemura1,Uemura2}.
This provides strong evidence for an unconventional pairing mechanism in the kagome superconductor CsV$_3$Sb$_5$.
 
 Moreover, we observe a clear difference in the temperature dependence of the anisotropies related to magnetic penetration depth and upper critical fields, again signaling a clear deviation from the Ginzburg-Landau theory. This situation finds a clear parallel with the data from the well-known 
 two-gap superconductor MgB$_2$ \cite{Angst_PRL_2002, Fletcher_PRL_2005}, Fe-based Sm- and Nd-1111 systems \cite{Weyeneth, Gupta}, 122 pnictide superconductors: Ba(Fe$_{1-x}$Co$_x$)$_2$As$_2$ \cite{Prozorov, Tanatar}, (Ba$_{1-x}$K$_x)$Fe$_2$As$_2$\cite{Khasanov_PRL_2009}; FeSe$_{0.5}$Te$_{0.5}$ \cite{Bendele}, CaKFe$_4$As$_4$ \cite{Rustem_PRB_2019}, etc., where two different temperature variations of $\gamma_{\lambda}(T)$ and $\gamma_{B_{c2}}(T)$ were attributed to multiband superconductivity. Thus, in comparison to well-established multigap superconductors, we find further support for the multigap behavior in this compound. Therefore, our results provide microscopic evidence of anisotropic multigap superconductivity in the kagome superconductor CsV$_3$Sb$_5$ and encourage further theoretical and experimental research on the kagome superconductors. 

Finally, ZF-$\mu$SR experiments suggest that across $T_{c}$, the relaxation rates do not change. This suggests that within the statistical accuracy, no clear signature of TRS breaking is observed in the superconducting state of CsV$_3$Sb$_5$.


\section{Methods} \label{sec:Experimental-Techniques}

Single crystals of CsV$_3$Sb$_5$ were grown from Cs ingots (purity 99.9\%), V 3-N powder (purity 99.9\%) and Sb grains (purity 99.999\%) using the self-flux method, similar to the growth of  RbV$_3$Sb$_5$ \cite{Yin}. The eutectic mixture of CsSb and CsSb$_2$ was mixed with VSb$_2$ to form a composition with approximately 50 at.\% Cs$_x$Sb$_y$ and 50 at.\% VSb$_2$. The mixture was put into an alumina crucible and sealed in a quartz ampoule under partial argon atmosphere. The sealed quartz ampoule was heated to 1273~K in 12~h and kept there for 24~h. Then it was cooled down to 1173~K at 50~K/h and further to 923~K at a slower rate. Finally, the ampoule was removed from the furnace and decanted with a centrifuge to separate the CsV$_3$Sb$_5$ crystals from the flux. The obtained crystals have a typical size of $4\times4\times1$~mm$^3$ and are stable in air over a period of at least several months. As shown in Fig.~1~(a) and (b), the flux-grown single crystals possess an obvious hexagonal symmetry, while the x-ray Laue diffraction images demonstrate the single crystallinity of the material. 

  The magnetization measurements were performed in a Quantum Design magnetic property measurement system SQUID magnetometer under field-cooled (FC) and zero-field-cooled (ZFC) conditions. The XRD pattern was collected using a Bruker D8 x-ray diffractometer with Cu K$_\alpha$ radiation ($\lambda$~=~0.15418~nm) at room temperature. Electrical transport measurements were carried out in a Quantum Design physical property measurement system (PPMS-14T). The longitudinal electrical resistivity was measured using a four-probe method with the current flowing in the $ab$ plane.
  
  We performed transverse-field (TF) ${\mu}$SR experiments using the Dolly spectrometer ($\pi E1$ beamline) at the Paul Scherrer Institute (Villigen, Switzerland). Since the crystals were rather thick ($\sim 1$~mm), they were mounted in a single layer using Apiezon N grease, to form a mosaic covering an area of $7\times7$~mm$^2$. The Dolly spectrometer is equipped with a standard veto setup, providing a low-background ${\mu}$SR signal. All TF- ${\mu}$SR experiments were done after field-cooling the sample with the applied field either along the kagome plane or perpendicular to it. The ${\mu}$SR time spectra were analyzed using the open software package \textsc{Musrfit} \cite{Suter_MuSRFit_2012}.
  
  {\bf Notes added}\\
While preparing this manuscript, we came to know about another ${\mu}$SR work studying the high temperature charge ordered phase of CsV$_3$Sb$_5$ [arXiv:2107.10714 (2021) (https://arxiv.org/abs/2107.10714)].

{\bf Acknowledgments}\\
${\mu}$SR experiments were performed at the Swiss Muon Source (S$\mu$S), Paul Scherrer Institute (PSI), Switzerland. The work of RG was supported by the Swiss National Science Foundation (SNF-Grant No.\ 200021-175935). H.C.L. was supported by National Natural Science Foundation of China (Grant No. 11822412 and 11774423), Ministry of Science and Technology of China (Grant No.\ 2018YFE0202600 and 2016YFA0300504) and the Beijing Natural Science Foundation (Grant No.\ Z200005).

{\bf Author contributions}\\
R.G., D.D., C.M., Z.G., R.K., and C.B. carried out ${\mu}$SR experiments. R.G., D.D., C.M., and R.K. performed ${\mu}$SR data analysis. Q.Y., Z.T., C.G., and H.C.L. provided and characterized samples. H.L. supervised the work at PSI. R.G., D.D., C.M., Z.G., and T.S. prepared the manuscript with notable inputs from all authors.

{\bf Competing interests}\\
The authors declare no competing interests.

{\bf Data availability:}\\
The data supporting the findings of this study are available within the paper and in the Supplementary Information. The raw data are available from the corresponding authors upon reasonable request.

\end{document}


\title{ Microscopic evidence of anisotropic multigap superconductivity \\ in the kagome superconductor CsV$_3$Sb$_5$}

\author{Ritu Gupta\footnotemark[2]}
 \email{ritu.gupta@psi.ch}
 
  \affiliation{Laboratory for Muon Spin Spectroscopy, Paul Scherrer Institute, CH-5232 Villigen PSI, Switzerland}

 \author{Debarchan Das}
\thanks{These authors contributed equally}
 \affiliation{Laboratory for Muon Spin Spectroscopy, Paul Scherrer Institute, CH-5232 Villigen PSI, Switzerland}
 
  \author{Charles Hillis Mielke III}
\thanks{These authors contributed equally}
 \affiliation{Laboratory for Muon Spin Spectroscopy, Paul Scherrer Institute, CH-5232 Villigen PSI, Switzerland}

   \author{Zurab Guguchia}
 \affiliation{Laboratory for Muon Spin Spectroscopy, Paul Scherrer Institute, CH-5232 Villigen PSI, Switzerland}
 
    \author{Toni~Shiroka}
 \affiliation{Laboratory for Muon Spin Spectroscopy, Paul Scherrer Institute, CH-5232 Villigen PSI, Switzerland}
  \affiliation{Laboratorium für Festk{\"o}rperphysik, ETH Z{\"u}rich, Z{\"u}rich, CH-8093, Switzerland}
  
    \author{Christopher Baines}
 \affiliation{Laboratory for Muon Spin Spectroscopy, Paul Scherrer Institute, CH-5232 Villigen PSI, Switzerland}

 \author{Marek Bartkowiak}
 \affiliation{Laboratory for Neutron and Muon Instrumentation, Paul Scherrer Institute, CH-5232 Villigen PSI, Switzerland}
 
    \author{Hubertus Luetkens}
 \affiliation{Laboratory for Muon Spin Spectroscopy, Paul Scherrer Institute, CH-5232 Villigen PSI, Switzerland}

\author{Rustem Khasanov}
 \email{rustem.khasanov@psi.ch}
 \affiliation{Laboratory for Muon Spin Spectroscopy, Paul Scherrer Institute, CH-5232 Villigen PSI, Switzerland}
\author{Qiangwei Yin}
\affiliation{Department of Physics and Beijing Key Laboratory of Opto-electronic Functional Materials \& Micro-nano Devices, Renmin University of China, Beijing 100872, China}
\author{Zhijun Tu}
\affiliation{Department of Physics and Beijing Key Laboratory of Opto-electronic Functional Materials \& Micro-nano Devices, Renmin University of China, Beijing 100872, China}
\author{Chunsheng Gong}
\affiliation{Department of Physics and Beijing Key Laboratory of Opto-electronic Functional Materials \& Micro-nano Devices, Renmin University of China, Beijing 100872, China}
\author{Hechang Lei}
\email{hlei@ruc.edu.cn}
\affiliation{Department of Physics and Beijing Key Laboratory of Opto-electronic Functional Materials \& Micro-nano Devices, Renmin University of China, Beijing 100872, China}

\date{\today}

\maketitle

\uppercase{\section*{Supplementary Information}}
\subsection{Electrical transport measurements}

The electrical resistivity was measured over the temperature interval 2 to 300 K and in magnetic fields up to 1.5 T using a standard AC four-probe technique implemented in a Quantum Design PPMS platform. In order to estimate the upper critical field ($\mu_0H_{\rm c2}$) values in two different field orientations, we measured the electrical resistivity with the field applied parallel to the $c$ axis and to $ab$ plane, respectively. Figures~S1~(a) and (b) show the temperature dependence of electrical resistivity with the applied field along these two directions. The upper critical fields $\mu_0H{\rm c2}$ versus $T_{c}$ are shown in Fig.~S1~(c), with $T_{c}$ defined at the zero value of resistivity. 

\begin{figure}[htb!]
\includegraphics[width=0.78\linewidth]{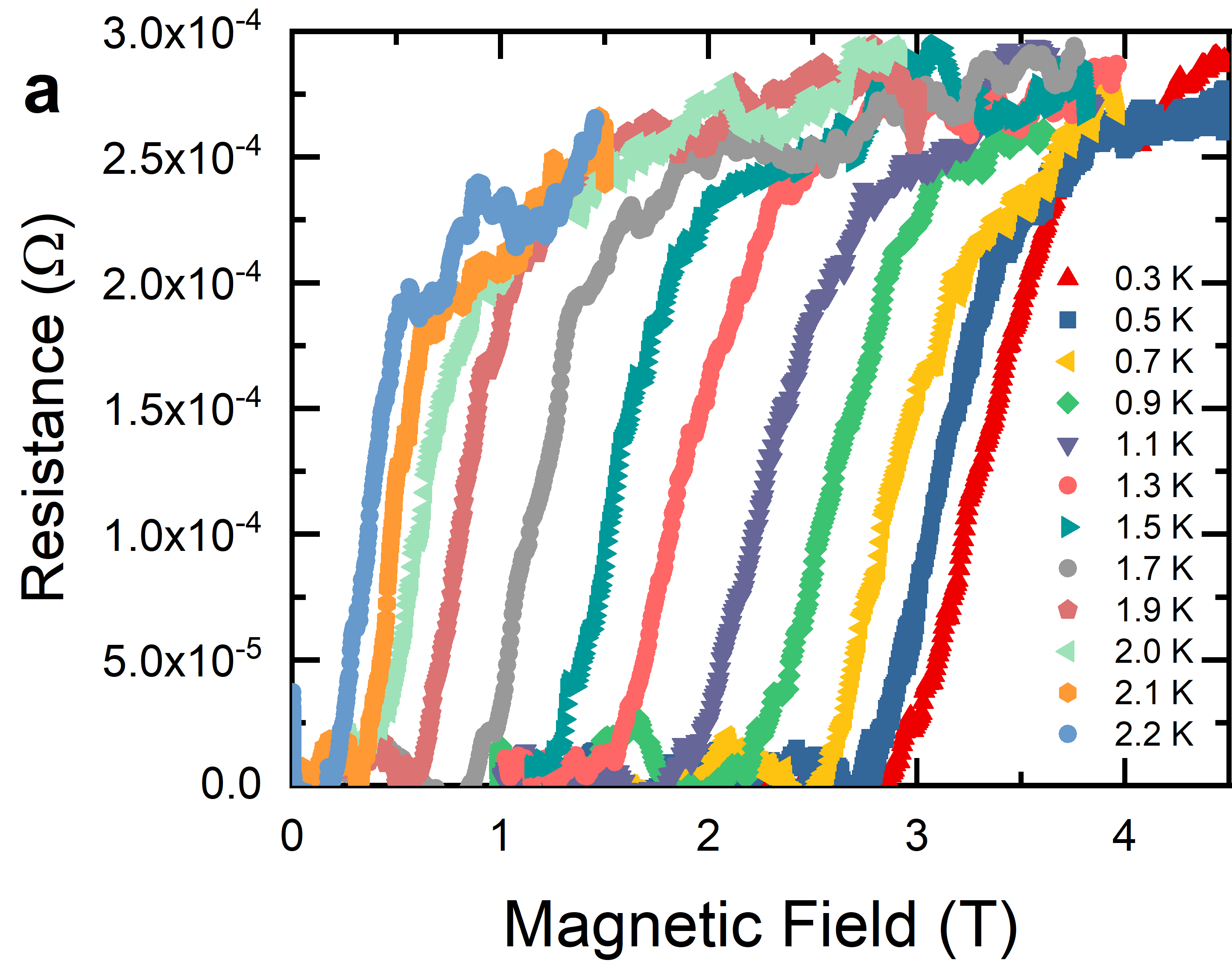}
\includegraphics[width=0.8\linewidth]{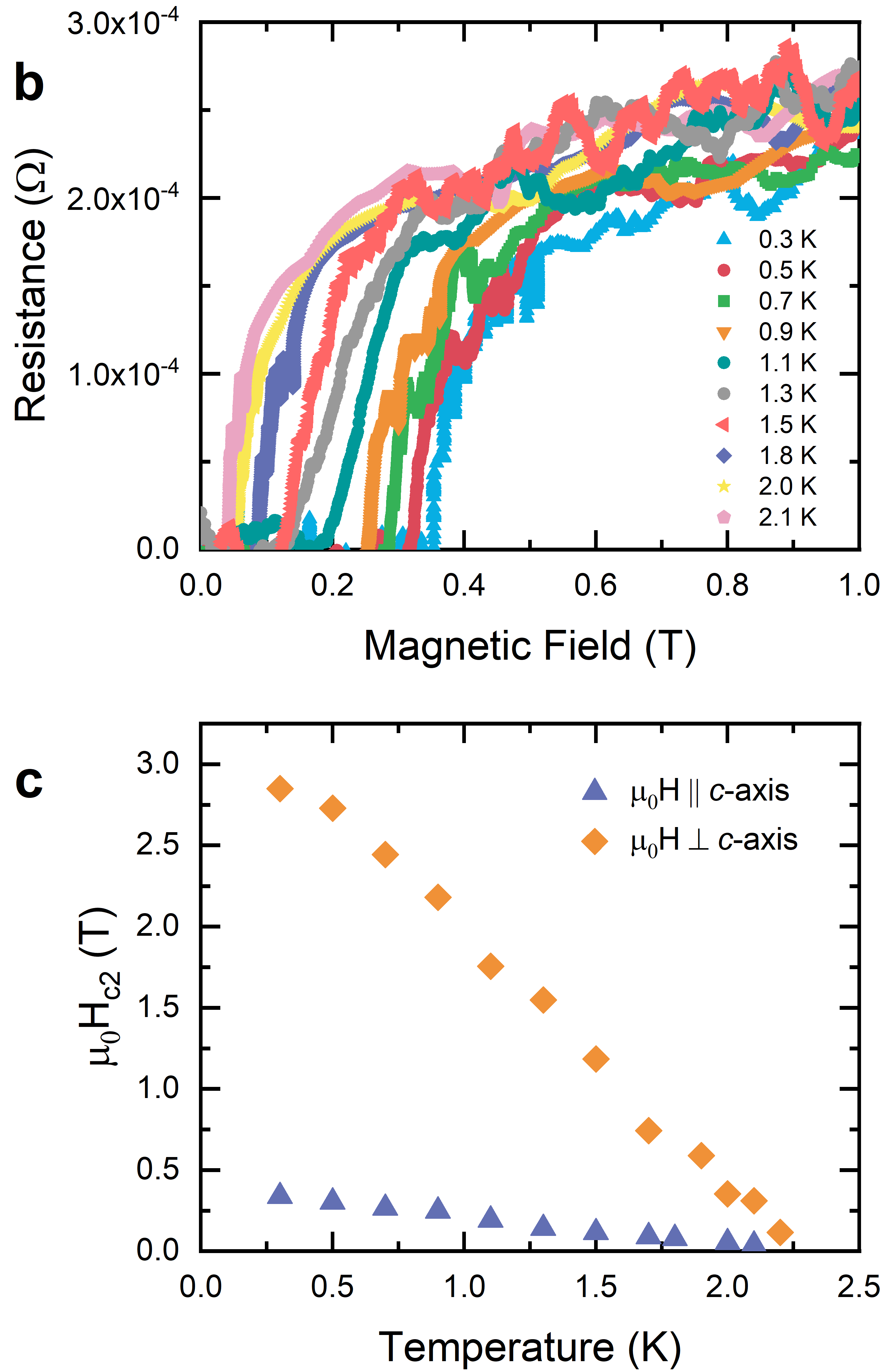}
\caption{Electrical resistivity of CsV$_3$Sb$_5$ as a function of temperature with applied field parallel to the $ab$ plane (a) and to the $c$ axis (b). Upper critical field vs $T_{c}$ phase diagram for both orientations (c).}
\label{fig:S1}
\end{figure}
\begin{figure}[htb!]
\includegraphics[width=0.85\linewidth]{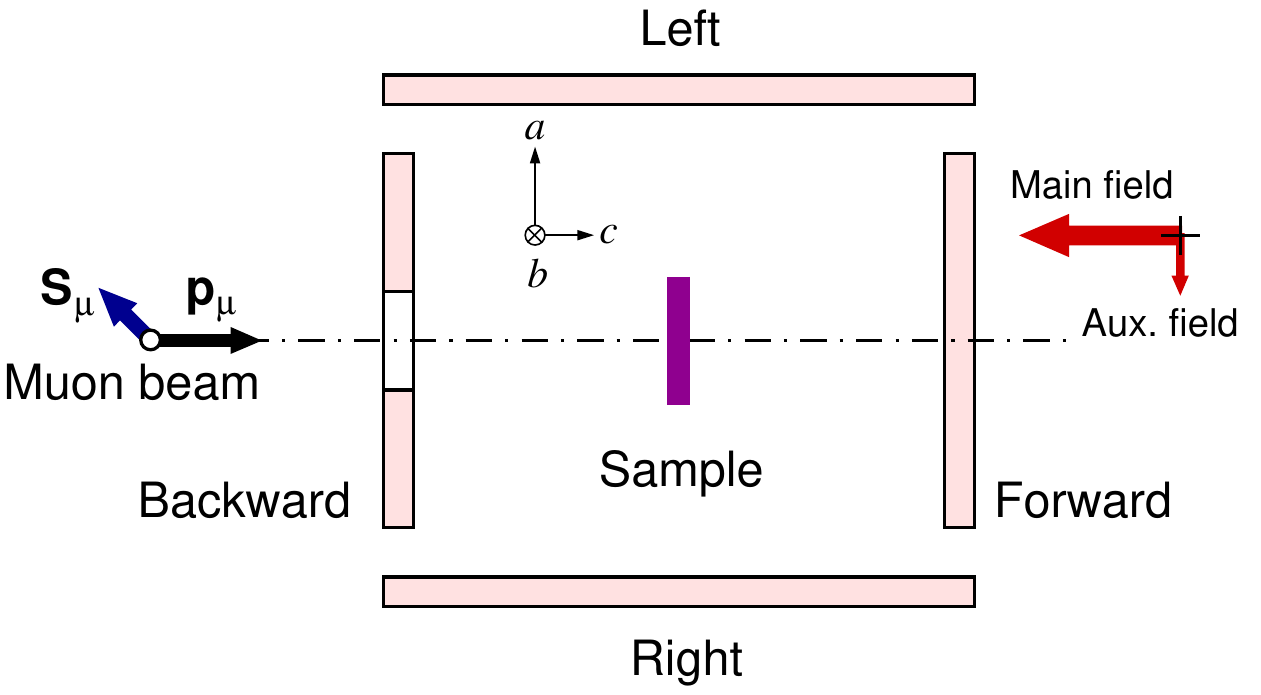}
\caption{Top view of the Dolly detectors along with the muon spin and momentum directions.}
\label{fig:S2}
\end{figure}

\subsection{Detector arrangement of \texorpdfstring{$\mu$SR}{muSR} spectrometer}

Figure S2 shows the top view of the positron detectors in the  Dolly $\mu$SR spectrometer, here grouped into two sets: Forward (F) - Backward (B) and Left (L) - Right (R). In the muon-spin rotated mode, $\boldsymbol{{S}_{\mu}}$ is made to rotate by an angle of $\sim~43^{\circ}$ with respect to its momentum $\boldsymbol{{S}_{\mu}}$ by means of crossed electric- and magnetic fields (Wien filter). In this configuration, for a  sample oriented with the crystallographic axes as shown in Fig.~S2 (as in our case), one component of the muon spin is parallel to the $ab-$plane, while the other is parallel to the crystallographic $c$ axis. Thus, by considering the positron counts in the F-B and L-R detectors separately, one can probe the zero-field response for $\boldsymbol{{S}_{\mu}}||~ ab$ and $\boldsymbol{{S}_{\mu}}||~ c$, respectively.

\subsection{Skewed Gaussian function}

The TF-$\mu$SR time spectra were analyzed by a skewed Gaussian (SKG) function, where the field distribution is given by \cite{Suter_Internal}:
\begin{equation*}
P\textsubscript{skg}(B)=\sqrt{\frac{2}{\pi}}\frac{\gamma_\mu}{\sigma_++\sigma_-}\begin{cases}
\textsubscript{exp}\big[-\frac{1}{2}\frac{(B-B_0)^2}{(\sigma_+/\gamma_\mu)^2}\big], \quad\!\!  B \geq B_0 \\
\textsubscript{exp}\big[-\frac{1}{2}\frac{(B-B_0)^2}{(\sigma_-/\gamma_\mu)^2}\big], \quad\!\! B < B_0
\end{cases}
\end{equation*}
Here, the maximum in the $P(B)$ distribution is taken as $B_0$, while $\gamma_\mu$~=~135.5~MHz/T represents the muon gyromagnetic ratio. We can determine the mean value of the first ($M_1$), second ($M_2$), and third ($M_3$) moments of the SKG distribution \cite{Suter_Internal}:
\begin{equation}
M_1 \equiv \left<B\right> = B_0-\frac{\sigma}{\gamma_\mu}\sqrt{\frac{2}{\pi}}(\zeta-1),
\end{equation}
\begin{equation}
M_2 = \bigg(\frac{\sigma}{\gamma_\mu}\bigg)^2\frac{1}{\pi}[\pi(1-\zeta+\zeta^2)-2(\zeta-1)^2],
\end{equation}

\begin{equation}
 M_3 = \bigg(\frac{\sigma}{\gamma_\mu}\bigg)^3\sqrt{\frac{2}{\pi^3}}(\zeta-1)[\pi(1-3\zeta+\zeta^2)-4(\zeta-1)^2],
\end{equation}
where $\sigma=\sigma_+$ and $\zeta=\sigma_-/\sigma_+$.
In order to fit the SKG distribution to the experimental TF-$\mu$SR
data, the transformation from the field domain to the time domain was performed via \cite{Suter_Internal}:
\begin{equation}
\mathrm{SKG}(t) = \int_{-\infty}^{+\infty}\!P_\mathrm{skg}(B)\cos(\gamma_\mu Bt)\,\mathrm{d}B.
\end{equation}

Thus, the TF-$\mu$SR asymmetry spectra were fitted to the equation:
\begin{equation}
A(t)=A_s\; {\rm SKG}(t)+A_{\rm BG}\;\cos(\gamma_\mu B_{\rm ext }+\phi).
\end{equation}
$A_s$ and $A_{\rm BG}$ are the initial asymmetries of the sample ($s$) and background (BG) contributions, respectively. From the fit, the values of the first, second, and third moments were determined. Close to $T_c$, the $P(B)$ distributions become nearly symmetric, so we used a single Gaussian instead of a skewed Gaussian. The temperature variation of $\sigma_+$ and $\sigma_-$ and $B_0$ for two field orientations is shown in Fig.~3.

\subsection{Anisotropy in magnetic penetration depth} 

In case of an anisotropic superconductor, the effective mass tensor is given by: \cite{Thiemann_PRB_1989}
\begin{equation}
m_{\rm eff}=\left(\begin{array}{ccc} m_i^* & 0 & 0\\ 0 & m_j^* & 0\\ 0 & 0 & m_k^* \end{array}\right).
\end{equation}
Here, $m_i^*$ represents the effective mass of the charge carriers flowing along the $i^\mathrm{th}$ principal axis. For a magnetic field applied along a principal axis of the effective mass tensor, the field penetration depth is given as:\cite{Thiemann_PRB_1989}
\begin{equation}
 \lambda_{jk}^{-2}=\frac{1}{\lambda_j \lambda_k}
 \label{eq:lambda_jk}
\end{equation}

By using Eq.~\ref{eq:lambda_jk}, the out-of-plane component of the magnetic penetration depth, $\lambda_c^{-2}$, could be obtained from $\lambda_{ab}^{-2}(T)$ and $\lambda_{ab,c}^{-2}(T)$ as: 
\begin{equation}
\lambda_{c}^{-2} = \frac{\lambda_{ab}^{2}}{\lambda_{ab,c}^{4}}.
 \label{eq:lambda_c}
 \end{equation}
Figure 3(a) and (b) in the main paper show the resulting temperature dependence of $\lambda_{ab}^{-2}$ and $\lambda_c^{-2}$.

\begin{figure}[htb!]
\includegraphics[width=0.8\columnwidth]{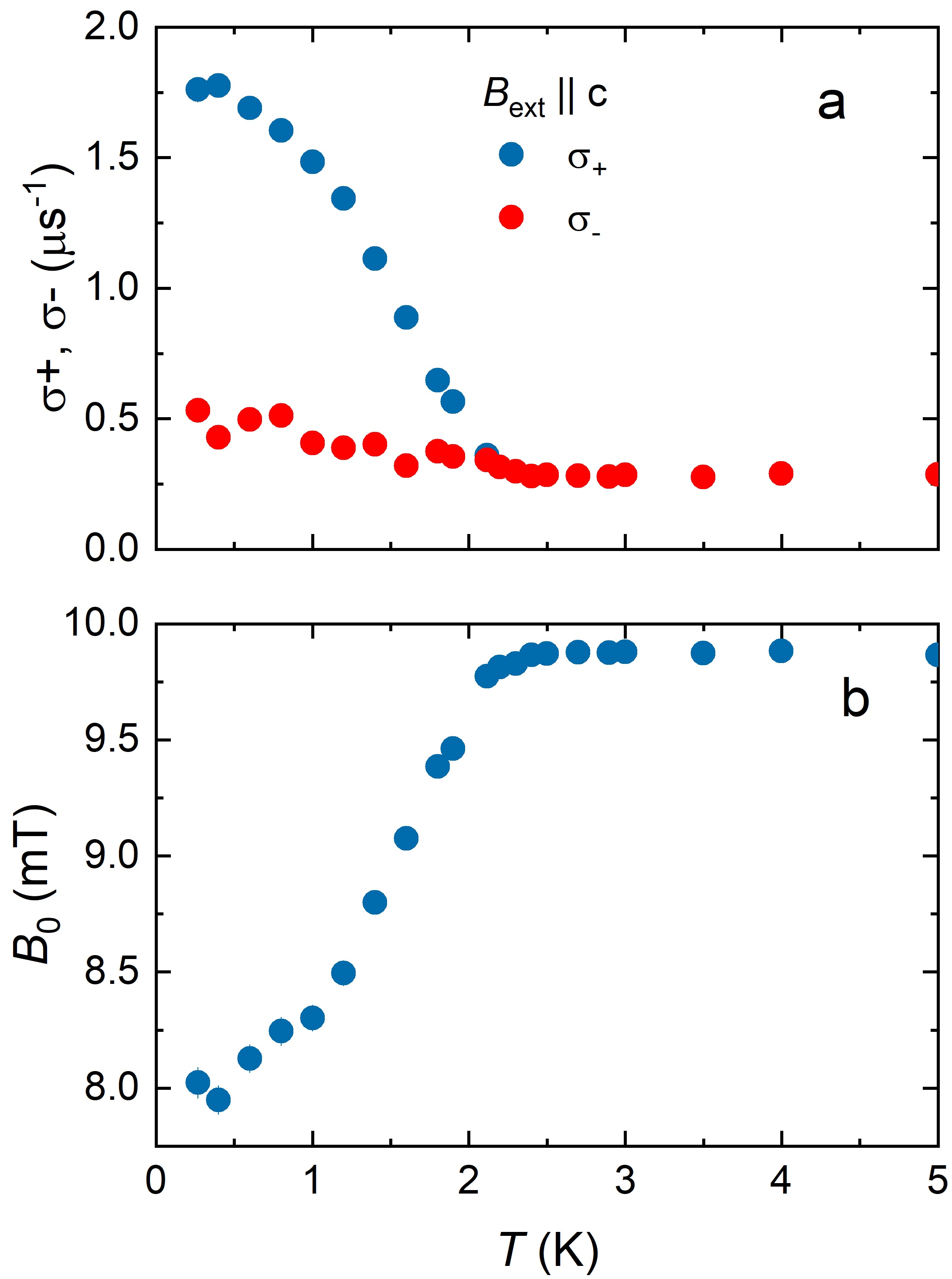}
\includegraphics[width=0.8\columnwidth]{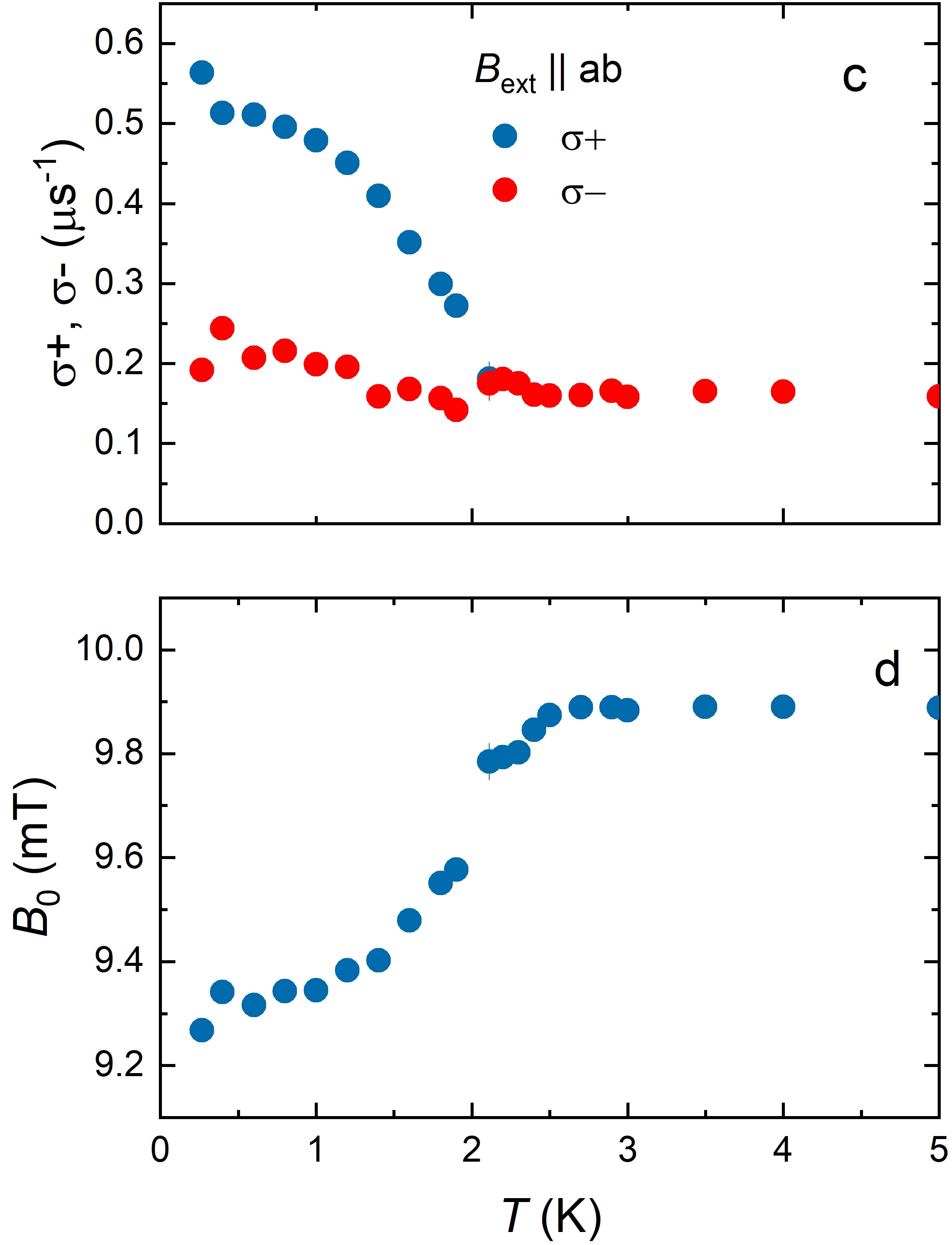}
\caption{Temperature variation of $\sigma_+$, $\sigma_-$, and $B_0$, as obtained by fitting the TF-$\mu$SR spectra by means of Eq.~(5). In panels (a) and (b) the field is parallel to  the $c$-axis; in (c) and (d) the field is parallel to the $ab$ plane.}
\label{fig:S3}
\end{figure}
\subsection{Zero Field analysis}
The zero field spectra collected for both directions above and below the superconducting transition were fitted by:
\begin{equation}
A(t)=A_0\exp(-\Lambda t)G\textsubscript{KT},
\end{equation}
where $G\textsubscript{KT}$ \cite{Kubo} is the Gaussian Kubo-Toyabe function \cite{Kubo},  which defines the muon spin depolarization rate associated with the nuclear magnetic moments and $\Lambda$ represents the electronic-spin relaxation rate. $G\textsubscript{KT}$ is defined as: 
\begin{equation}
G\textsubscript{KT}(t)=\frac{1}{3}+\frac{2}{3}(1-\sigma\textsubscript{ZF}^2t^2)\exp\bigg(-\frac{\sigma\textsubscript{ZF}^2t^2}{2}\bigg),
\end{equation}
where $\sigma\textsubscript{ZF}$ is the width of the nuclear dipolar field distribution experienced by the muon-spin ensemble.
